\documentclass[pra,twocolumn,superscriptaddress,notitlepage,showpacs,showkeys]{revtex4-1}
\usepackage{graphicx,amsmath,amsfonts,amssymb,upgreek,txfonts,color}
\usepackage[colorlinks,linkcolor=blue,citecolor=blue,urlcolor=blue,breaklinks=true]{hyperref}
\usepackage[utf8x]{inputenc} 

\begin{document}

\title{Dynamics of a hybrid optomechanical system in the framework of the generalized linear response theory}

\author{B. Askari} 
\address{Department of Physics, Shahid Beheshti University, Tehran 19839, Iran}

\author{A. Dalafi} 
\email{a\_dalafi@sbu.ac.ir}
\address{Laser and Plasma Research Institute, Shahid Beheshti University, Tehran 19839-69411, Iran}

\date{\today}

\begin{abstract}
We present a theoretical study of the linear response of a driven-dissipative hybrid optomechanical system consisting of an interacting one-dimensional Bose–Einstein condensate (BEC) to an external time-dependent perturbation in the framework of the generalized linear response theory. Using the equations of motion of the open quantum system Green's function, we obtain the linear responses of the optical and atomic modes of the hybrid system and show how the atom-atom interaction of the BEC atoms affects the two normal resonances of the system as well as the anti-resonance frequency at which the optical field amplitude of the cavity becomes zero. Furthermore, an interpretation of the anti-resonance phenomenon is presented based on the the optical spectral density and the self-energy.

\end{abstract}


\maketitle
\section{Introduction}
During the recent two decades, the science of quantum optomechanics, which deals with the radiation pressure coupling of a cavity optical field to the vibrational mode of a mechanical oscillator, has been developing very quickly both in the theoretical and experimental aspects \cite{Aspelmeyer,Bowen book}. Optomechanical systems (OMSs) have had a significant contribution in many applications like: displacement and force sensing \cite{Kipp Vaha 2007,Tsa X 2012,Wimm 2014,aliNJP,Fani2020,aliDCEBECforce,aliAVSbook2020}, ground-state cooling of the vibrational modes of a mechanical oscillator \cite{Teufel}, synchronization of the mechanical oscillators \cite{Mari}, and generation of entanglement \cite{Palom}. Another kind of OMSs can be formed by an optical cavity consisting of a Bose-Einstein condensate (BEC) or an ensemble of ultra-cold atoms where the fluctuation of the collective excitation of the atomic field behaves like an effective mechanical mode \cite{Kanamoto,Brenn Science,Ritter Appl. Phys. B}. In such hybrid OMSs the radiation pressure of the cavity field can be coupled with the collective atomic mode as well as the vibrational mode of a mechanical oscillator \cite{Dalafi Dispersive,Dalafi CK,Dalafi 2BEC,Dalafi EQC}.

In many interesting quantum optical phenomena like normal mode splitting \cite{Dobr,Grob}, electromagnetically induced transparency (EIT) \cite{Imamoglu} or optomechanically induced transparency (OMIT) \cite{OMIT1,OMIT2,oMITReview2018} it is important to study the response of the system to an external time-dependent perturbation. In such cases the response of the system to the external source can be obtained by the standard liner response theory (SLRT) which is based on a closed model of the quantum system being initially in contact with a thermal bath at a finite temperature \cite{Coleman,Flensberg}. The deficiency of the SLRT is that the effect of dissipation has to be entered into the theory phenomenologically since the environment which is responsible for fluctuation an dissipation is not modeled within the SLRT \cite{Stefanucci}.

Recently, several researches have been conducted to generalize the SLRT to the theory of open quantum systems both in the Schr{\"o}dinger picture through the master equation approach \cite{greenScarlatella1,banPRA17,shenPRA17} and in the Heisenberg picture through the quantum Langevin equations (QLEs) \cite{shenOptLett18,aliGreen}. In the so-called generalized linear response theory (GLRT) the linear response of an open quantum system to an external time-dependent perturbation is investigated while the effect of the environment is taken into account in the mathematical modeling as a multi-mode quantum field with an infinite number of degrees of freedom which interacts with the open quantum system. In this way, the GLRT has been enriched enough mathematically to predict the dissipative phenomena by itself in spite of the SLRT which needs some phenomenological manipulations to be able to describe the dissipative effects.

Interestingly, the GLRT predicts a set of equations of motion for the open quantum system Green's functions which are derived in the Hisenberg picture through the QLEs \cite{aliGreen}. The linear response of the open quantum system to the external source can be calculated through the solutions of Green's functions equations of motion. It is worth mentioning again that the effect of dissipation is taken into consideration in the GLRT without necessity of any phenomenological manipulation and it is the superiority of GLRT over the SLRT. As a practical application, in Ref.\cite{aliGreen} the linear response of a standard bare OMS to an external time-dependent perturbation has been studied in the framework of GLRT.

In the present article, we are going to investigate the linear responses of the optical and atomic modes of a hybrid OMS consisting of an interacting cigar-shaped Bose-Einstein condensate (BEC) trapped inside an optical cavity to a time-dependent perturbation while both the optical and the atomic modes are investigated as open quantum systems. It is shown that the system behaves as a standard OMS with a difference that the hybrid system has an extra interaction term (in addition to the optomechanical interaction ) which is due to inter-atomic collisions of the BEC atoms. Using the equations of motions of the open quantum system Green's functions predicted by the GLRT, we obtain the linear responses of the system to a weak probe laser which acts as a time-dependent perturbation and show how the atom-atom interaction affects the optical and atomic responses of the system.

One of the most interesting features of the present hybrid OMS is that it behaves as a system consisting of of two coupled quantum oscillators which one of them (the Bogoliubov mode of the BEC) functions as an atomic parametric amplifier through the atom-atom interaction of the BEC atoms which is the atomic analog of the optical parametric amplifier \cite{dalafi1,dalafi2,dalafi3,dalafi4}. The study of the linear responses of such hybrid OMS shows that it has two resonance frequencies corresponding to the two normal modes of the system and an anti-resonance frequency \cite{Belbasi,Joe} which occurs for the optical mode since it is the oscillator which is driven directly by the time-dependent perturbation. It is demonstrated how the position of the anti-resonance frequency can be manipulated by the \textit{s}-wave scattering frequency of BEC atoms which itself is controllable through the transverse trapping frequency. Furthermore, it is also shown that the amount of splitting between the normal modes can be controlled by the coupling laser pumping rate which can change the effective optomechanical coupling between the optical and atomic modes. Finally, an interpretation of the optical response behavior especially the manifestation of the anti-resonance phenomenon is presented based on the optical spectral density and self-energy.

The paper is structured as follows: In Sec. \ref{Hamiltonian} the system and its Hamiltonian are introduced. In Sec. \ref{dynamics} the dynamics of the system is described using the theory of open quantum systems, and in Sec. \ref{GLRT} the responses of the optical and atomic modes to the time-dependent perturbation are studied in the framework of GLRT and the effects of atom-atom interaction to the system response is investigated. Finally, our conclusions are summarized in Sec. \ref{Conclusions}.

\section{System Hamiltonian}\label{Hamiltonian}
Consider a hybrid OMS formed by an optical cavity with length $L$ and resonance frequency $ \omega_{0} $ consisting of a BEC of $N$ two-level atoms with mass $m_{a}$ and transition frequency $\omega_{a}$ confined in a cylindrically symmetric trap with a transverse trapping frequency $\omega_{\mathrm{\perp}}$ and negligible longitudinal confinement along the $x$ direction. The cavity is driven by a coupling laser with frequency $ \omega_{c} $  and wave number $k=\omega_{c}/c$ at the rate of $\eta_c=\sqrt{2\mathcal{P}\kappa/\hbar\omega_{c}}$ through one of its mirrors where $\mathcal{P}$ is the laser power and $\kappa$ is the cavity decay rate.

In the dispersive regime, where the coupling laser frequency is far detuned from the atomic resonance so that $\Delta_{a}=\omega_{c}-\omega_{a}\gg\Gamma_{a}$ with $\Gamma_{a}$ being the atomic linewidth, the dynamics of atoms can be described within an effective 1D model by quantizing the atomic motional degree of freedom along the cavity axis, x \cite{Masch Ritch 2004,Dom JB,Nagy Ritsch 2009}. Therefore, the system Hamiltonian in the frame rotating at the coupling laser frequency is given by
\begin{eqnarray}\label{H1}
H&=&\int_{-L/2}^{L/2} dx \Psi^{\dagger}(x)\Big[\frac{-\hbar^{2}}{2m_{a}}\frac{d^{2}}{dx^{2}}+\hbar U_{0} \cos^2(kx) a^{\dagger} a\nonumber\\
&&+\frac{1}{2} U_{s}\Psi^{\dagger}(x)\Psi(x)\Big] \Psi(x)-\hbar\Delta_{c} a^{\dagger} a + i\hbar\eta(a - a^{\dagger}),\nonumber\\
\end{eqnarray}
where $ \Psi(x) $  is the quantum field operator of the atoms in the framework of the second quantization formalism, and $ a $ is the annihilation operator of the single optical mode of the cavity. Besides, $ \Delta_{c}=\omega_{c}-\omega_{0} $ is the detuning of the coupling laser from the cavity resonance, $U_{0}=g_{0}^{2}/\Delta_{a}$ is the optical lattice barrier height per photon with $g_{0}$ being the vacuum Rabi frequency, $U_{s}=\frac{4\pi\hbar^{2} a_{s}}{m_{a}}$ and $a_{s}$ is the two-body \textit{s}-wave scattering length \cite{Masch Ritch 2004,Dom JB}. The last terms in the Hamiltonian of Eq.(\ref{H1}) denotes the effect of the coupling laser which drives the optical mode of the cavity.

On the other hand, if the average number of the cavity photons is low enough so that the condition $U_{0}\langle a^{\dagger}a\rangle\leq 10\omega_{R}$ is satisfied with $\omega_{R}=\frac{\hslash k^{2}}{2m_{a}}$ being the recoil frequency of the condensate atoms, and under the Bogoliubov approximation \cite{Nagy Ritsch 2009}, the atomic field operator can be approximated by the following single-mode quantum field \cite{dalafi1}
\begin{equation}\label{opaf}
\Psi(x)=\sqrt{\frac{N}{L}}+\sqrt{\frac{2}{L}}\cos(2kx) c.
\end{equation}
Here, the first term, corresponding to the so-called condensate mode, has been considered as a c-number in the Bogoliubov approximation and the operator $ c $ in the second term (the so-called Bogoliubov mode) corresponds to the quantum fluctuations of the atomic field about the classical condensate mode. By substituting the atomic field operator of Eq.(\ref{opaf}) into the Hamiltonian of Eq.(\ref{H1}), the system Hamiltonian is simplified as
\begin{eqnarray}\label{Hc}
H&=&\hbar\delta_{c} a^{\dagger} a + i\hbar\eta(a-a^{\dagger})+\hbar\Omega_{c} c^{\dagger}c+\hbar\zeta a^{\dagger}a (c+c^{\dagger})\label{subH}\nonumber\\
&&+\frac{1}{4}\hbar\omega_{sw}(c^{2}+c^{\dagger 2}),
\end{eqnarray}
where $ \delta_{c}=-\Delta_{c}+\frac{1}{2}N U_{0} $ is the Stark-shifted cavity frequency due to the presence of the BEC, and $ \Omega_{c}=4\omega_{R}+\omega_{sw} $ is the frequency of the Bogoliubov mode. The important point that should be emphasized is that the resonance frequency of the cavity has been shifted from $\omega_0$ to $\tilde{\omega}_0=\omega_0 +\frac{1}{2}N U_{0}$ due to the presence of the BEC. In this way, the hybrid system behaves effectively as a bare optomechanical cavity with the shifted optical frequency $\tilde{\omega}_0$ which interacts with an effective mechanical oscillator whose role is played by the Bogoliubov mode of the BEC through a radiation pressure interaction with the effective optomechanical coupling $ \zeta=\frac{\sqrt{2N}}{4}U_{0} $ \Big(the fourth term in Eq.(\ref{Hc})\Big).

The last term which behaves as an atomic parametric amplifier \cite{dalafi1,dalafi2,dalafi3,dalafi4} corresponds to the atom-atom interaction with $ \omega_{sw}=8\pi\hbar a_{s}N/m_{a}Lw^2 $ being the \textit{s}-wave scattering frequency of the atomic collisions ($ w $ is the waist radius of the optical mode). The important point is that the strength of the atom-atom interaction which is determined by the \textit{s}-wave scattering frequency can be controlled experimentally by manipulating the transverse trapping frequency $\omega_{\mathrm{\perp}}$ which can change the waist radius of the optical mode $ w $ \cite{Morsch}.

\section{Dynamics of the system}\label{dynamics}
The dynamics of the hybrid OMS described by the Hamiltonian of Eq.(\ref{subH}) is fully characterized by the following set of nonlinear QLEs in the framework of open quantum systems \cite{Zubairy,Gardiner,Carmichael,Bowen book} 
\begin{subequations}
\begin{eqnarray}
\dot{a}&=&-(i\delta_{c}+\kappa/2)a-i\zeta a(c+c^{\dagger})-\eta+\sqrt{\kappa}\delta a_{in},\label{NLQLEa}\\
\dot{c}&=&-(i\Omega_{c}+\gamma/2)c-\frac{i}{2}\omega_{sw}c^{\dagger}-i\zeta a^{\dagger}a+\sqrt{\gamma}\delta c_{in},\label{NLQLEc}
\end{eqnarray}
\end{subequations}
where both the optical mode of the cavity and the Bogoliubov mode of the BEC are affected by their corresponding reservoirs. As is seen from QLEs (\ref{NLQLEa}-\ref{NLQLEc}), the system dynamics is affected by two uncorrelated quantum noise sources. 

(i) The optical input vacuum noise $\delta a_{in}$ arising from all the optical modes outside the cavity satisfying the Markovian correlation functions, i.e., $\langle\delta a_{in}(t)\delta a_{in}^{\dagger}(t^{\prime})\rangle=(n_{ph}+1)\delta(t-t^{\prime})$, $\langle\delta a_{in}^{\dagger}(t)\delta a_{in}(t^{\prime})\rangle=n_{ph}\delta(t-t^{\prime})$ with the average thermal photon number $n_{ph}$ which is nearly zero at optical frequencies \cite{Zubairy,Gardiner,Carmichael,Bowen book}. (ii) The atomic quantum noise operator $\delta c_{in}$ arising from the harmonic trapping potential in which the BEC has been confined and also from the extra modes of the BEC which have been neglected in the single-mode approximation of Eq.(\ref{opaf}) as has been discussed in Refs.\cite{dalafi4,Zhang2010}. All of these extra atomic modes behave as an atomic reservoir which not only injects the quantum noise $\delta c_{in}$ into the atomic system but also make the Bogoliubov mode of the BEC dissipate at the damping rate of $\gamma$. Besides, the atomic quantum noise also satisfies the same Markovian correlation functions as those of the optical noise \cite{Zhang2010}

The nonlinear QLEs(\ref{NLQLEa}-\ref{NLQLEc}) can be linearized by expanding the quantum operators around their respective classical mean values as $ a=\alpha+\delta a $ and $ c=\beta+\delta c $ where $ \delta a $ and $ \delta c $ are small quantum fluctuations around the mean fields $ \alpha $ and $ \beta $ whose steady-states are obtained as
\begin{subequations}
\begin{eqnarray}
\alpha&=&-\frac{\eta}{i\Delta+\kappa/2},\\
\beta&=&-\zeta|\alpha|^{2}\frac{\Omega^{(-)}+i\gamma/2}{\Omega^{(+)}\Omega^{(-)}+\gamma^{2}/4},
\end{eqnarray}
\end{subequations}
where $ \Delta=\delta_{c}+2\beta_{R}\zeta $ with $ \beta_{R} $ being the real part of the complex mean field $ \beta $, and $ \Omega^{(\pm)}=\Omega_{c}\pm\frac{1}{2}\omega_{sw} $ while the quantum fluctuations together with their Hermitian conjugates satisfy the following linearized set of ordinary differential equations
\begin{equation}\label{nA}
\dot{\boldsymbol{u}}(t)=\boldsymbol{\chi}_{0} \boldsymbol{u}(t)+\boldsymbol{u}_{in}(t),
\end{equation}
where $\boldsymbol{u}(t)=[\delta a(t),\delta a^{\dagger}(t),\delta c(t),\delta c^{\dagger}(t)]^{T}$ is the vector of continuous variable fluctuation operators,
$ \boldsymbol{u}_{in}(t)=[\sqrt{\kappa}\delta a_{in},\sqrt{\kappa}\delta a^{\dagger}_{in},\sqrt{\gamma}\delta c_{in},\sqrt{\gamma}\delta c^{\dagger}_{in}]^{T} $ is the corresponding vector of quantum noises, and also
\begin{equation}\label{chi0}
\boldsymbol{\chi}_{0}=\left(\begin{array}{cccc}
-\frac{\kappa}{2}-i\Delta & 0 & -i\alpha\zeta & -i\alpha\zeta \\
   0 & -\frac{\kappa}{2}+i\Delta & i\alpha^{\ast}\zeta & i\alpha^{\ast}\zeta \\
    -i\alpha^{\ast}\zeta & -i\alpha\zeta & -\frac{\gamma}{2}-i\Omega_{c} & -\frac{i}{2}\omega_{sw} \\
    i\alpha^{\ast}\zeta & i\alpha\zeta & \frac{i}{2}\omega_{sw} & -\frac{\gamma}{2}+i\Omega_{c}\\
    \end{array}\right),
\end{equation}
is the drift matrix. It is worth reminding that the solutions to Eq.(\ref{nA}) are stable only if all the eigenvalues of the matrix $ \boldsymbol{\chi}_{0} $ have negative real parts. The stability conditions can be obtained, for example, by using the Routh-Hurwitz criteria \cite{RH}.

Now, the linearized QLEs of the system can be solved by taking the Fourier transform of Eq.(\ref{nA}) as
\begin{equation}\label{nAw}
\boldsymbol{u}(\omega)=\boldsymbol{\chi}(\omega)\boldsymbol{u}_{in}(\omega),
\end{equation}
where the susceptibility matrix $\boldsymbol{\chi}(\omega)$ is obtained as
\begin{equation}\label{chiw}
\boldsymbol{\chi}(\omega)=\Big(-i\omega\boldsymbol{1}-\boldsymbol{\chi}_{0}\Big)^{-1},
\end{equation}
where $\boldsymbol{1}$ is the $4\times 4$ identity matrix. It is obvious that the Fourier components of both the optical and the atomic quantum fluctuations can be obtained from Eq.(\ref{nAw}) in terms of the Fourier components of the quantum noises of the system.

\section{Linear response of hybrid OMS in the framework of GLRT}\label{GLRT}
In this section we investigate the linear response of the hybrid OMS described in Secs.\ref{Hamiltonian} and \ref{dynamics} to a weak external time-dependent perturbation which is executed by a weak probe laser with frequency $\omega_{p}$ which drives the cavity with the rate $\eta_{p}$ whose absolute value is much lower than that of the coupling laser. Therefore, the dynamics of the system is affected by the following time-dependent perturbation in the frame rotating at the coupling laser frequency
\begin{equation}\label{Vt}
 V(t)=\hbar\eta_{p}\delta a e^{i\omega_{pc}t}+\hbar\eta_{p}^\ast\delta a^\dagger e^{-i\omega_{pc}t},
\end{equation}
where $\omega_{pc}=\omega_{p}-\omega_{c}$ is the detuning between the probe and coupling lasers frequencies 

In Ref.\cite{aliGreen}, the linear response of a standard OMS has been investigated in the framework of the GLRT. On the other hand, in Secs.\ref{Hamiltonian} and \ref{dynamics}, it was shown that a hybrid OMS consisting of an atomic BEC behaves effectively as a standard OMS. In this way, the response of the optical and atomic field fluctuations of the hybrid OMS to the external time-dependent perturbation can be obtained based on the GLRT as follows 
\begin{subequations}
	\begin{eqnarray}
	\langle\delta a(t)\rangle=\langle\delta a\rangle_0+\eta_{p}\int_{-\infty}^{+\infty} dt^{\prime} G_{aa}^R(t-t^{\prime}) &&e^{i\omega_{pc}t^{\prime}}\nonumber\\
	+\eta_{p}^\ast\int_{-\infty}^{+\infty} dt^{\prime} G_{aa^\dag}^R(t-t^{\prime}) e^{-i\omega_{pc}t^{\prime}},\label{mat}\\
	\langle\delta c(t)\rangle=\langle\delta c\rangle_0+\eta_{p}\int_{-\infty}^{+\infty} dt^{\prime} G_{ca}^R(t-t^{\prime}) &&e^{i\omega_{pc}t^{\prime}}\nonumber\\
	+\eta_{p}^\ast\int_{-\infty}^{+\infty} dt^{\prime} G_{ca^\dag}^R(t-t^{\prime}) e^{-i\omega_{pc}t^{\prime}},\label{mct}
	\end{eqnarray}
\end{subequations}
where $\langle\delta a\rangle_0=0$ and $\langle\delta b\rangle_0=0$ are the steady-state mean values of the optical and atomic field fluctuations in the absence of the time-dependent perturbation and the system retarded Green's functions have been defined as
\begin{subequations}
	\begin{eqnarray}
	G_{aa}^R(t)&=&-i\theta(t)\langle [\delta a(t),\delta a(0)]\rangle_0,\label{Gaa}\\
	G_{aa^\dag}^R(t)&=&-i\theta(t)\langle [\delta a(t),\delta a^\dag(0)]\rangle_0\label{Gaad},\\
	G_{ca}^R(t)&=&-i\theta(t)\langle [\delta c(t),\delta a(0)]\rangle_0,\label{Gca}\\
	G_{ca^\dag}^R(t)&=&-i\theta(t)\langle [\delta c(t),\delta a^\dag(0)]\label{Gcad}\rangle_0.
	\end{eqnarray}
\end{subequations}
The first two Green's functions of Eqs.(\ref{Gaa}),(\ref{Gaad}) are related to the optical field while Eqs.(\ref{Gca}),(\ref{Gcad}) represent the atomic field Green's functions. It should be noted that the time evolutions of all the operators in Eqs.(\ref{Gaa}-\ref{Gcad}) are obtained from the QLEs given by Eq.(\ref{nA}) derived in Sec.\ref{dynamics} and the subscript 0 means that all the expectation values should be calculated in the steady-state of the system in the absence of the perturbation. On the other hand, Eqs.(\ref{mat}) and (\ref{mct}) can be rewritten as 
\begin{subequations}
	\begin{eqnarray}
		\langle\delta a(t)\rangle&=&\eta_{p}^\ast \tilde G_{aa^\dag}^R(\omega_{pc}) e^{-i\omega_{pc}t}+\eta_{p} \tilde G_{aa}^R(-\omega_{pc}) e^{i\omega_{pc}t}.\\
		\langle\delta c(t)\rangle&=&\eta_{p}^\ast \tilde G_{ca^\dag}^R(\omega_{pc}) e^{-i\omega_{pc}t}+\eta_{p} \tilde G_{ca}^R(-\omega_{pc}) e^{i\omega_{pc}t},
	\end{eqnarray}
\end{subequations}
in terms of the definition of the Fourier transform of the Green's function, i.e., $\tilde{G}(\omega)=\int_{-\infty}^{+\infty}d\tau G(\tau) e^{i\omega\tau}$.

Since $\langle a(t)\rangle=\alpha+\langle\delta a(t)\rangle$, and $\langle c(t)\rangle=\beta+\langle\delta c(t)\rangle$ are the expectation values of the optical and atomic fields in the rotating frame, the responses of the optical and atomic fields to the time-dependent perturbation in the laboratory frame are obtained as
\begin{subequations}
	\begin{eqnarray}
	\langle a(t)\rangle &=& \alpha e^{-i\omega_{c}t}+\eta_{p}^\ast \tilde G_{aa^\dag}^R(\omega_{pc}) e^{-i(\omega_{c}+\omega_{pc})t}\nonumber\\
	&&+\eta_{p} \tilde G_{aa}^R(-\omega_{pc}) e^{-i(\omega_{c}-\omega_{pc})t},\label{Ra}\\
	\langle c(t)\rangle&=&\beta +\eta_{p}^\ast \tilde G_{ca^\dag}^R(\omega_{pc}) e^{-i\omega_{pc}t}\nonumber\\
	&&+\eta_{p} \tilde G_{ca}^R(-\omega_{pc}) e^{i\omega_{pc}t}.\label{Rc}
	\end{eqnarray}
\end{subequations}
As is seen from Eq.(\ref{Ra}), the optical mode has a central band oscillating with $\omega_{c}$ and two sidebands, the so-called Stokes and anti-Stokes sidebands, oscillating with $\omega_{c}\pm\omega_{pc}$. For the atomic mode the situation is the same with the difference that the central mode, i.e., the mean field $\beta$ in Eq.(\ref{Rc}), has no oscillation in laboratory frame.

Based on the GLRT, the Green's functions of an open quantum system satisfy a system of ordinary differential equations which are derived through the linearized QLEs, i.e., Eq.(\ref{nA}). and are called the Green's functions equations of motion which are given by the following compact equations
\begin{subequations}
	\begin{eqnarray}
	&& \frac{d}{dt} \boldsymbol{G}^R_{a^\dag}(t)= -i\delta(t)\boldsymbol{V}_{a^\dag}+\boldsymbol{\chi}_0 \boldsymbol{G}^R_{a^\dag}(t),\label{GAd}\\
	&& \frac{d}{dt} \boldsymbol{G}^R_{a}(t)= +i\delta(t)\boldsymbol{V}_{a}+\boldsymbol{\chi}_0 \boldsymbol{G}^R_{a}(t),\label{GA}
	\end{eqnarray}
\end{subequations}
where $ \boldsymbol{G}^R_{a^\dag}(t)= \Big(G^R_{aa^\dag}(t), G^R_{a^\dag a^\dag}(t), G^R_{ca^\dag}(t), G^R_{c^\dag a^\dag}(t)\Big)^{\rm T} $, $ \boldsymbol{G}^R_{a}(t)= \Big(G^R_{aa}(t), G^R_{a^\dag a}(t), G^R_{ca}(t), G^R_{c^\dag a}(t)\Big)^{\rm T} $, and $ \boldsymbol{V}_{a^\dag}:= (1,0,0,0)^{\rm T} $ and $ \boldsymbol{V}_{a}:= (0,1,0,0)^{\rm T} $ are fixed four-dimensional vectors. Now, by taking the Fourier transforms of Eqs.(\ref{GAd}) and (\ref{GA}) the Fourier components of the Green's function vectors are obtained as
\begin{subequations}
	\begin{eqnarray}
	&&  \boldsymbol{\tilde G}^R_{a^\dag}(\omega)= -i \boldsymbol{\chi}(\omega) \boldsymbol{V}_{a^\dag},\label{GAdw}\\
	&&  \boldsymbol{\tilde G}^R_{a}(\omega)= +i \boldsymbol{\chi}(\omega) \boldsymbol{V}_{a},\label{GAw},
	\end{eqnarray}
\end{subequations}
where $\boldsymbol{\chi(\omega)}$ is the susceptibility matrix defined by Eq.~(\ref{chiw}). As is seen from Eqs.~(\ref{GAdw}) and (\ref{GAw}), the system Green's functions in the frequency space, i.e., the Fourier transforms of Eqs.~(\ref{Gaa}-\ref{Gcad}) are obtained as
\begin{subequations}
	\begin{eqnarray} \label{greenparametric}
	&& \tilde G^R_{aa}(\omega)=+ i\chi_{aa^\dag} (\omega),\label{Gaa(w)}\\
	&& \tilde G^R_{aa^\dag}(\omega)=- i\chi_{aa} (\omega),\label{Gaad(w)}\\
	&& \tilde G^R_{ca}(\omega)=+ i\chi_{ca^\dag} (\omega),\label{Gca(w)}\\
	&& \tilde G^R_{ca^\dag}(\omega)=- i\chi_{ca} (\omega).\label{Gbad(w)}
	\end{eqnarray}
\end{subequations}

\subsection{the effects of atomic collisions and coupling laser on the system responses}
In this subsection, we investigate how the atom-atom interaction as well as the pumping rate of the coupling laser affect the linear responses of the optical and atomic modes of the hybrid OMS to the time dependent perturbation. Based on the linearized QLEs given by Eq.(\ref{nA}), it is obvious that the optical mode oscillates with frequency $\Delta$ while it can be easily shown that the atomic mode oscillates effectively with following frequency
\begin{equation}\label{wm}
\omega_{m}=\sqrt{\Big(4\omega_{R}+\frac{1}{2}\omega_{sw}\Big) \Big(4\omega_{R}+\frac{3}{2}\omega_{sw}\Big)},
\end{equation}
which is called the effective mechanical frequency of the Bogoliubov mode of the BEC \cite{dalafi5}. In the following we will study the dynamics of the hybrid system in the red detuned regime of $\Delta=\omega_m$ which is possible by fixing the coupling laser frequency at $\omega_c=\tilde{\omega}_0-\omega_m$. Obviously, under this condition the effective frequencies of the optical and atomic modes are equal.

Furthermore, we present our results based on the experimentally feasible parameters given in \cite{Ritter Appl. Phys. B, Brenn Science}. For this purpose, we consider a cavity with length $ L=178 \mu$m, damping rate of $\kappa=10^5$Hz, and bare frequency $ \omega_{0}=2.41494\times 10^{15} $Hz corresponding to a wavelength of $ \lambda=780 $nm which contains  $ N=5\times 10^5 $ Rb atoms. The atomic $ D_{2} $ transition corresponding to the atomic transition frequency $ \omega_{a}=2.41419\times 10^{15} $Hz couples to the mentioned mode of the cavity. The atom-field coupling strength is $ g_{0}=2\pi\times 14.1 $MHz and the recoil frequency of the atoms is $ \omega_{R}=23.7 $KHz. Furthermore, we assume that the equilibrium temperature of the BEC is $ T=0.1\mu $K, and the damping rate of the Bogoliubov mode of the BEC is $\gamma=10^{-4}\kappa$.

\begin{widetext}
	
	\begin{figure}
		\centering
		\includegraphics[width=7cm]{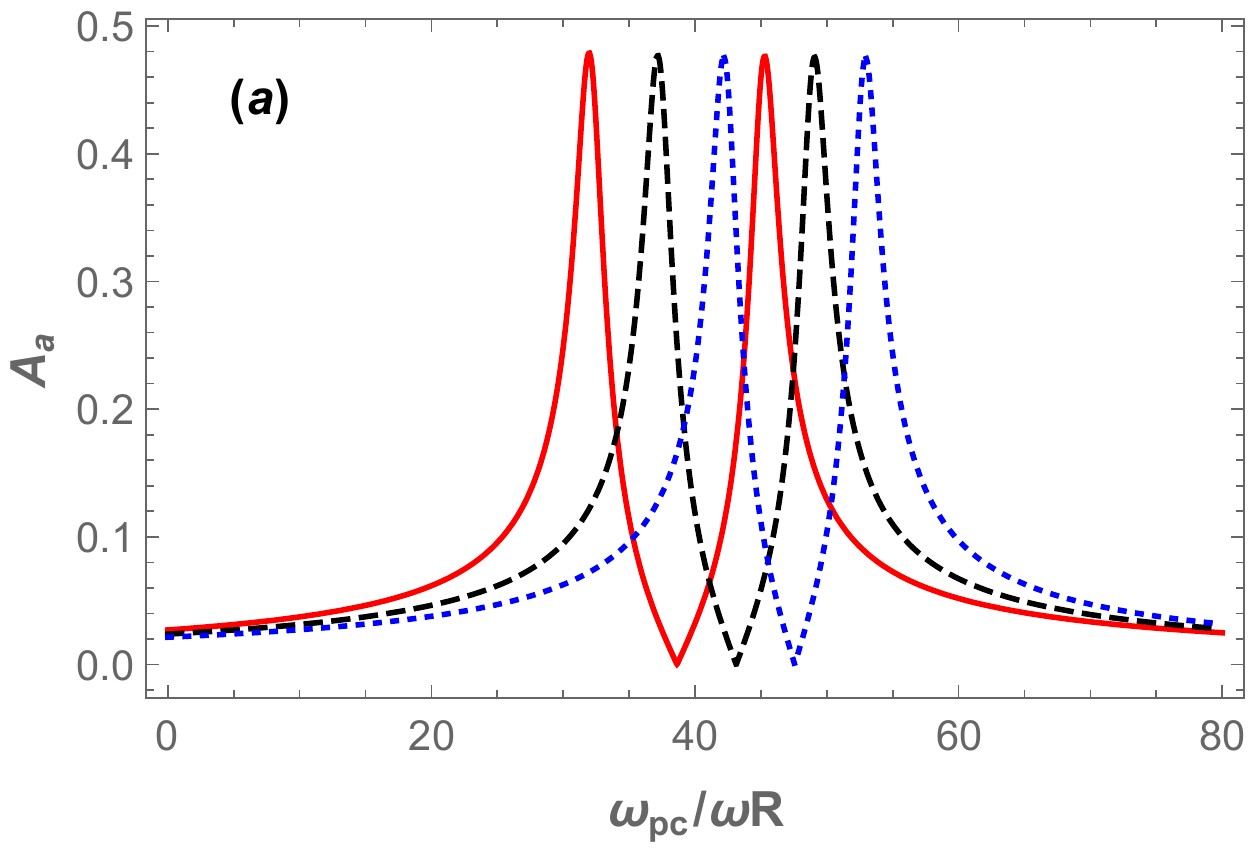}
		\includegraphics[width=7cm]{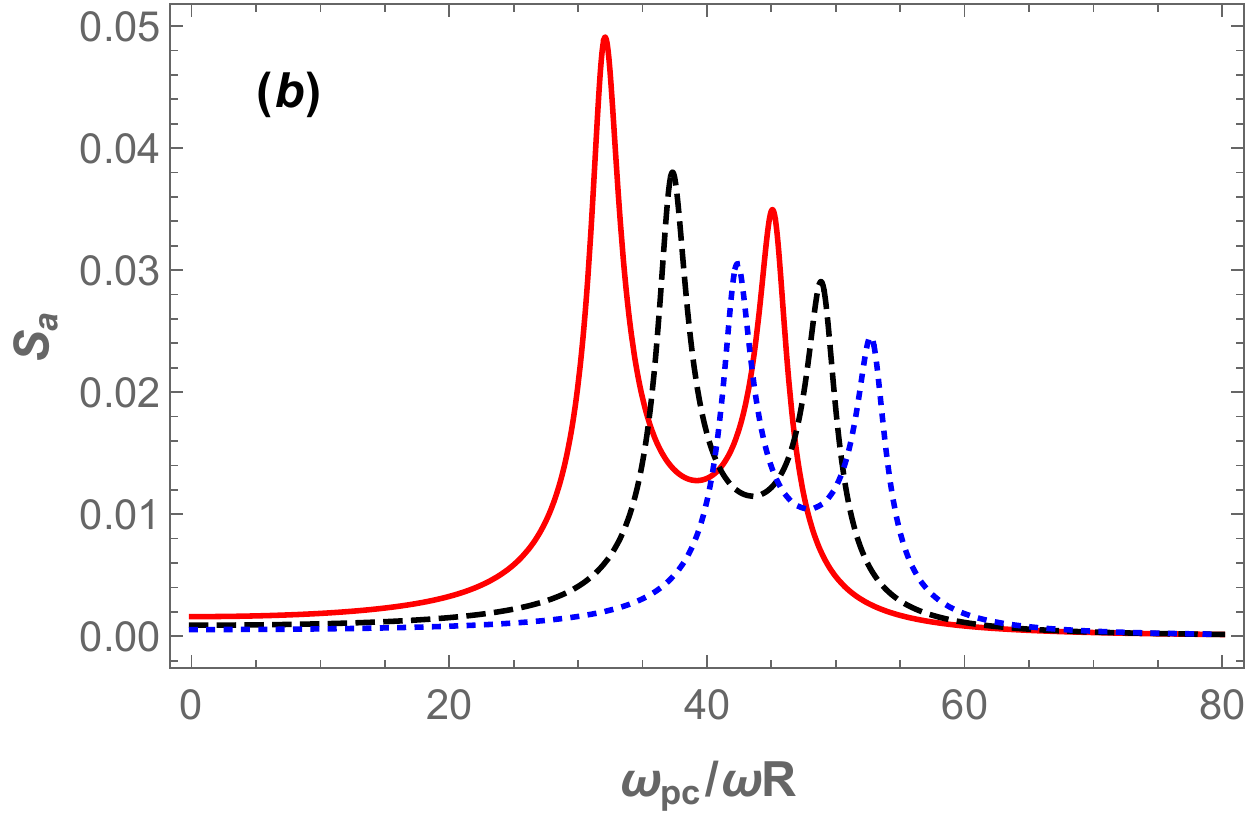}
		\includegraphics[width=7cm]{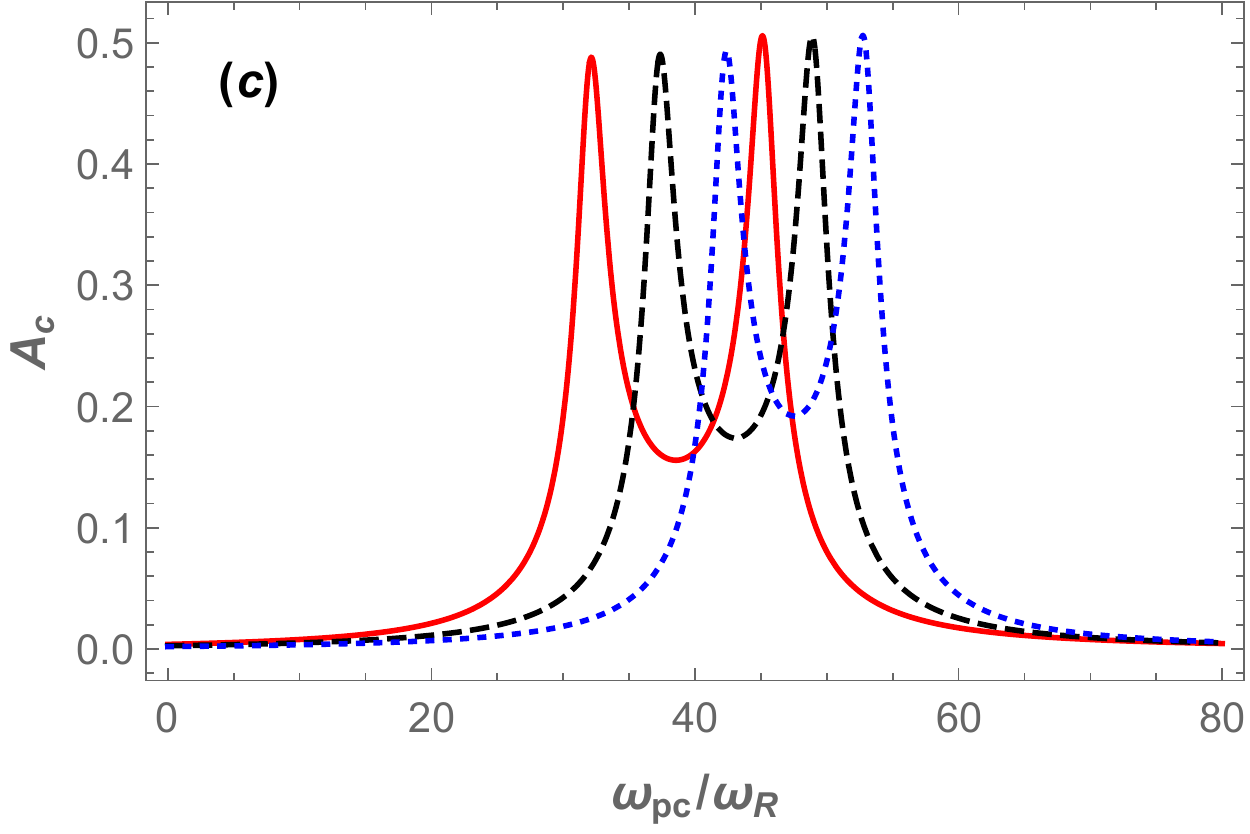}
		\includegraphics[width=7cm]{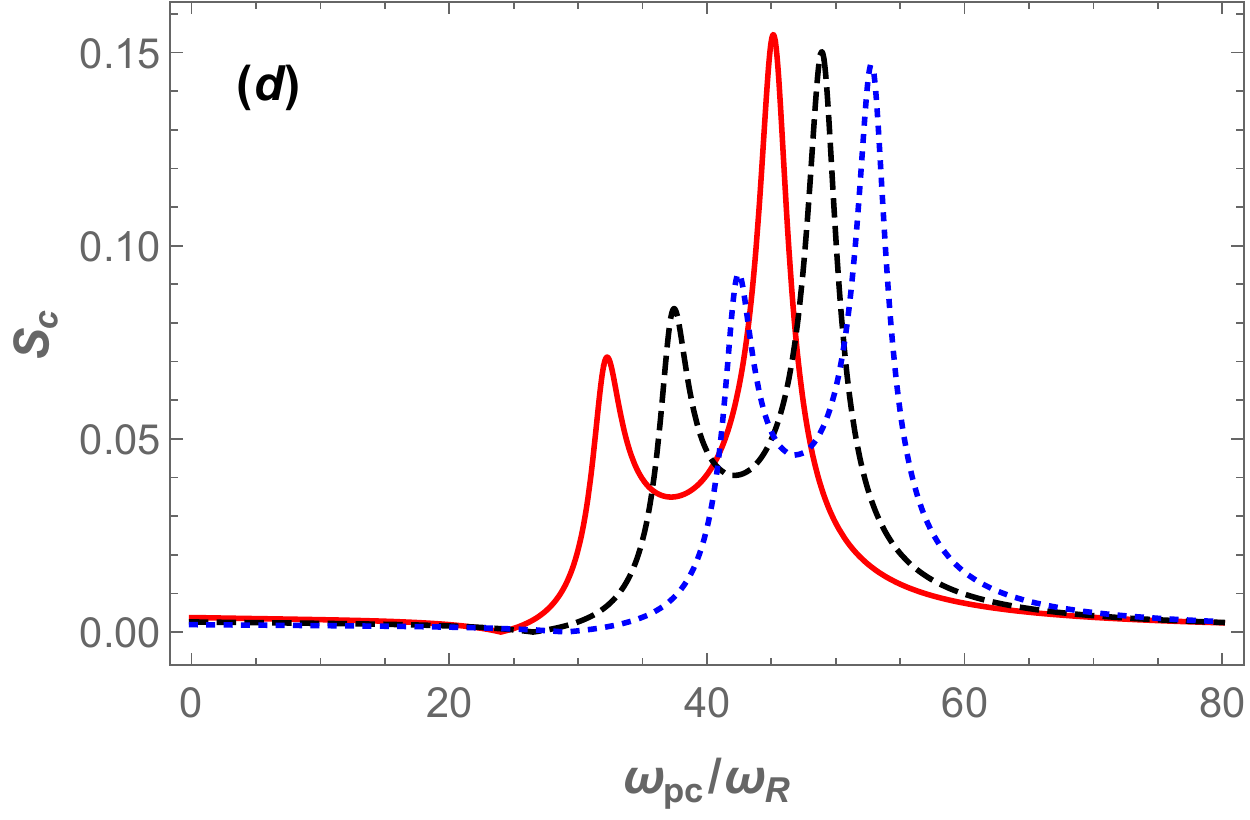}
		\caption{
			(Color online) (a),(b) The normalized amplitudes of the anti-Stokes  $A_a=\omega_{R}|\tilde G^R_{aa^\dag}(\omega_{pc}/\omega_{R})|$ and the Stokes $S_a=\omega_{R} |\tilde G^R_{aa}(-\omega_{pc}/\omega_{R})|$ sidebands of the optical field, and (c), (d) the anti-Stokes  $A_c=\omega_{R}|\tilde G^R_{ca^\dag}(\omega_{pc}/\omega_{R})|$ and the Stokes $S_c=\omega_{R} |\tilde G^R_{ca}(-\omega_{pc}/\omega_{R})|$ sidebands of the Bogoliubov mode of the BEC for three different values of the \textit{s}-wave scattering frequency: $\omega_{sw}=40\omega_R$ (red solid curve), $\omega_{sw}=45\omega_R$ (black dashed curve), and $\omega_{sw}=50\omega_R$ (blue dotted curve). It has been assumed that the system is in the red detuned regime of $\Delta=\omega_m$ while the cavity is driven at the rate of $\eta_c=05\kappa$ by the coupling laser.}
		\label{fig1}
	\end{figure}
	
\end{widetext}

In order to study the effect of atomic collisions on the responses of the optical and atomic modes of the system to the external time-dependent perturbation, in Fig.\ref{fig1} we have plotted the normalized amplitudes of the anti-Stokes  $A_a=\omega_{R}|\tilde G^R_{aa^\dag}(\omega_{pc}/\omega_{R})|$ \Big(Fig.\ref{fig1}(a)\Big) and the Stokes $S_a=\omega_{R} |\tilde G^R_{aa}(-\omega_{pc}/\omega_{R})|$  \Big(Fig.\ref{fig1}(b)\Big) sidebands of the optical field as well as  the anti-Stokes  $A_c=\omega_{R}|\tilde G^R_{ca^\dag}(\omega_{pc}/\omega_{R})|$ \Big(Fig.\ref{fig1}(c)\Big) and the Stokes $S_c=\omega_{R} |\tilde G^R_{ca}(-\omega_{pc}/\omega_{R})|$ \Big(Fig.\ref{fig1}(d)\Big) sidebands of the Bogoliubov mode of the BEC for three different values of the \textit{s}-wave scattering frequency: $\omega_{sw}=40\omega_R$ (red solid curve), $\omega_{sw}=45\omega_R$ (black dashed curve), and $\omega_{sw}=50\omega_R$ (blue dotted curve) while the cavity is driven at the rate of $\eta_c=0.5\kappa$.

All the results of Fig.\ref{fig1} have been obtained in the red detuned regime of  $\Delta=\omega_{m}$ where the coupling laser frequency is smaller than the effective frequency of the cavity as $\omega_c\approx\tilde{\omega}_0-\omega_m$. The condition $\Delta=\omega_{m}$ leads to an algebraic equation of order three for $\omega_c$ which can be solved for any fixed value of $\omega_{sw}$. Based on our numerical calculations, for at most one of the solutions of $\omega_c$, the system is stable based on the the Routh-Hurwitz criteria \cite{RH}. As is well-known, in the red detuned regime of optomechanics the anti-Stokes sideband is enhanced  while the Stokes sideband is attenuated \cite{Bowen book,Aspelmeyer}. That is why the Stokes amplitudes are much smaller than the anti-Stokes ones for both the optical and the atomic modes in Fig.\ref{fig1}. It means that in the red detuned regime both the optical and the atomic modes oscillate effectively with the anti-Stokes amplitude.

As is seen from Fig.\ref{fig1}, for each value of the \textit{s}-wave scattering frequency the anti-Stokes and Stokes amplitudes of the optical and atomic fields exhibit two peaks corresponding to the two resonances at the normal frequencies of the hybrid OMS due to the coupling between the optical and atomic modes since for $\eta_c=0.5\kappa$ the system is in the normal mode splitting regime \cite{Dobr,Grob} where the enhanced effective optomechanical, i.e., $\zeta |\alpha|$, is greater than the damping rate of the cavity. The other important point is that for each value of the \textit{s}-wave scattering frequency there is an anti-resonance between the two resonances of the normal modes which occurs at $\omega_{pc}=\omega_{m}$, which is equivalent to $\omega_{p}\approx\tilde{\omega}_{0}$, where the anti-Stokes amplitude of the optical mode becomes zero \Big(Fig.\ref{fig1} (a)\Big) while the anti-Stokes amplitude of the atomic field reduces to a nonzero minimum \Big(Fig.\ref{fig1} (c)\Big).

It is well-known that for a driven dissipative system consisting of two coupled oscillators there are two resonance frequencies corresponding to the two normal modes where the oscillation amplitudes of the two oscillators becomes very large. On the other hand as has been investigated in Refs.\cite{Belbasi,Joe}, there is an anti-resonance frequency between the two resonance frequencies where the oscillation amplitude of the oscillator that is directly driven by the external source becomes zero in the limit where its damping rate is much larger than that of the other oscillator. The anti-resonance phenomenon occurs because the phase of the first oscillator suffers a sudden change and becomes out of phase with the second one so that the motion of the first oscillator is quenched effectively by the second one \cite{Belbasi,Joe}.

The present hybrid OMS investigated in this article, is a quantum simulation of the above-mentioned classical system of coupled oscillators. The single-mode optical field of the cavity that is driven by the probe laser plays the role of the first oscillator which has been coupled to the Bogoliubov mode of the BEC (the second oscillator) through a radiation pressure interaction which is triggered by the coupling laser. That is why the phenomenon of anti-resonance occurs just for the optical mode of the cavity since it is the oscillator that is directly driven by an external source. Furthermore, in the present hybrid OMS $\kappa\gg\gamma$, which means that the damping rate of the first oscillator is much larger than the second one's.

The most important result that has been demonstrated in Fig.\ref{fig1} is the fact that the anti-resonance frequency is shifted to higher values as the \textit{s}-wave scattering frequency increases. Since the mechanical frequency of the Bogoliubov mode of the BEC, i.e., $\omega_m$ given by Eq.(\ref{wm}), increases by increasing the \textit{s}-wave scattering frequency and as the anti-resonance frequency occurs at $\omega_{pc}=\omega_m$, the position of anti-resonance is shifted to higher values by increasing $\omega_{sw}$. However, for an experimental observation of a pattern like that predicted theoretically in Fig.\ref{fig1} (a), one needs to have an estimation of the value of $\omega_{sw}$ which can be controlled by the transverse frequency of the optical trap \cite{Morsch} and can be measured through the  phase noise power spectrum of the cavity output field \cite{dalafi5} or, alternatively, via a Feshbach resonance by the application of an appropriate magnetic field \cite{Marte,Donley}.

\begin{figure}
	\centering
	\includegraphics[width=7cm]{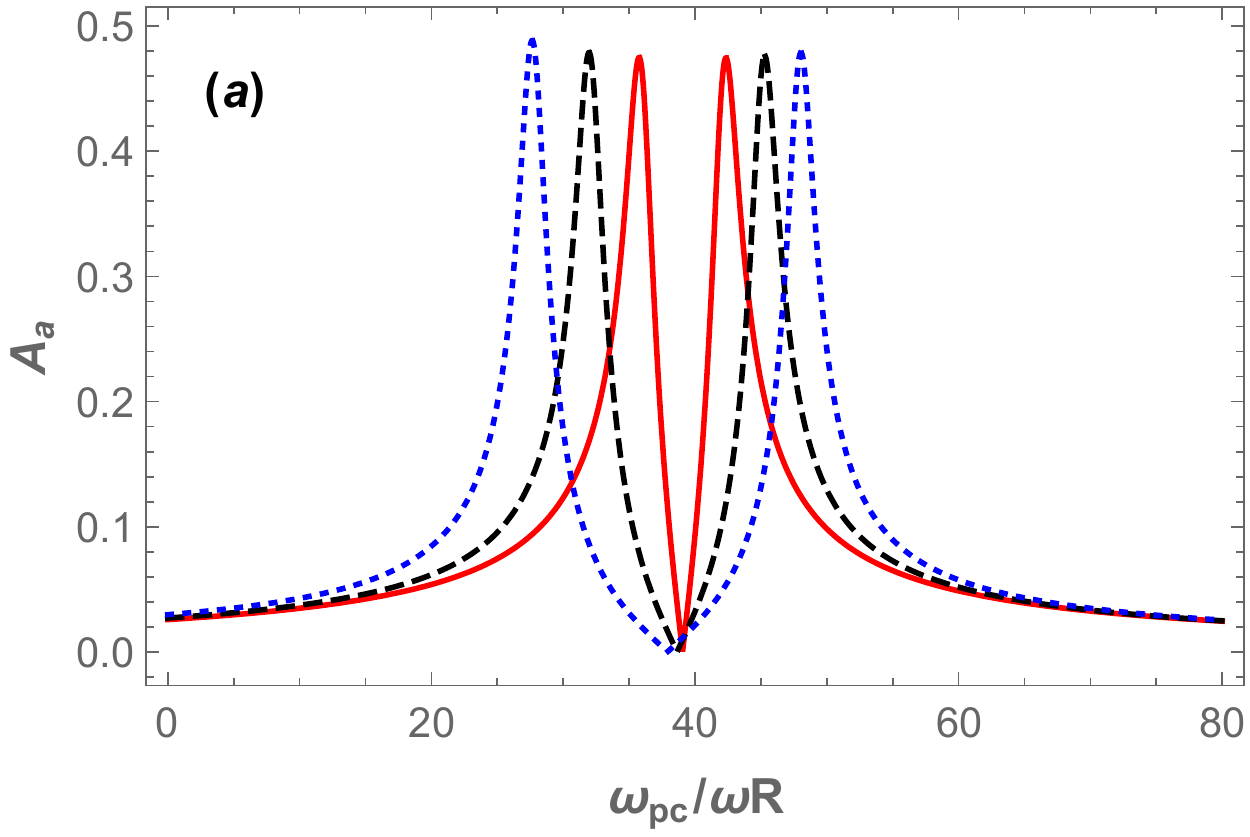}
	\includegraphics[width=7cm]{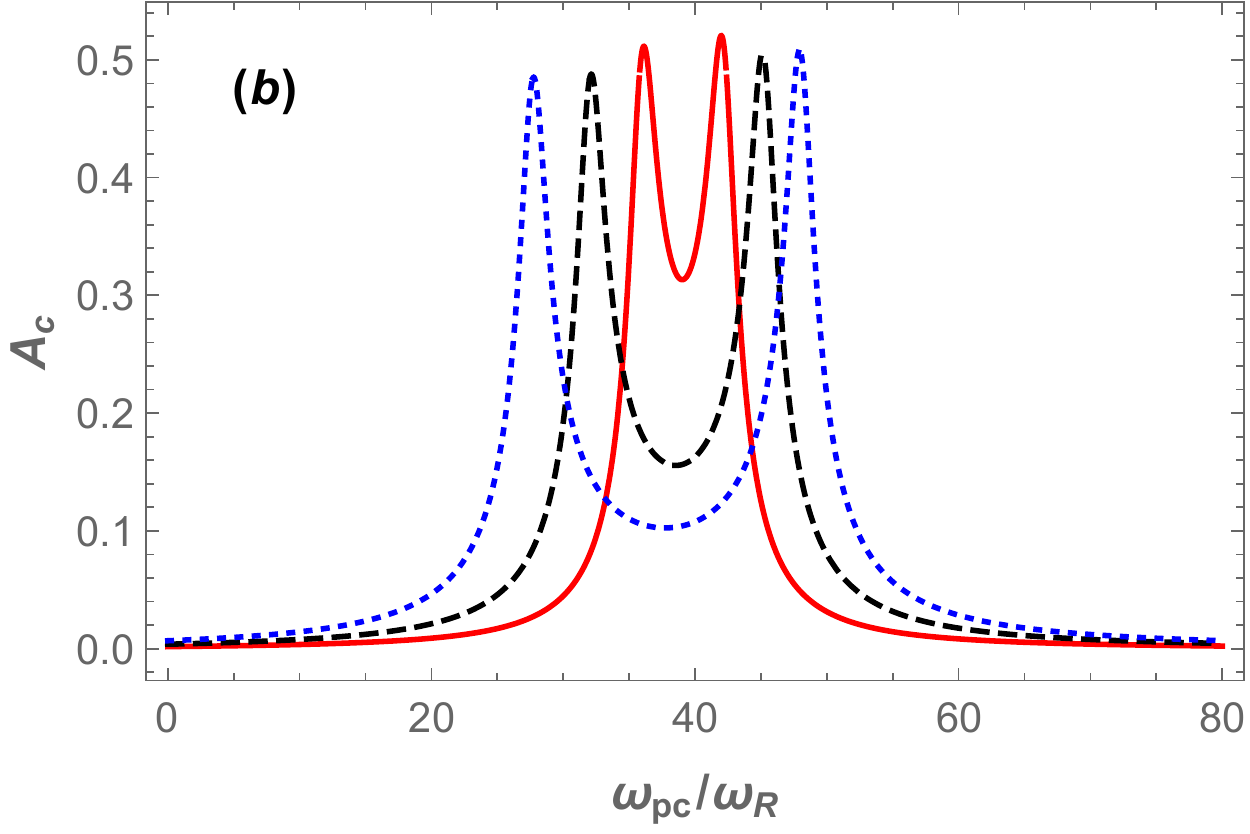}
	\caption{
		(Color online) (a) The normalized amplitudes of the anti-Stokes sideband  $A_a=\omega_{R}|\tilde G^R_{aa^\dag}(\omega_{pc}/\omega_{R})|$ of the optical field, and (b) the anti-Stokes sideband  $A_c=\omega_{R}|\tilde G^R_{ca^\dag}(\omega_{pc}/\omega_{R})|$ of the Bogoliubov mode of the BEC for three different values of the coupling laser pumping rate: $\eta_c=2.5\kappa$ (red solid curve),  $\eta_c=0.5\kappa$ (black dashed curve), and $\eta_c=7.5\kappa$ (blue dotted curve). It has been assumed that $\omega_{sw}=40\omega_{R}$ and the system is in the red detuned regime of $\Delta=\omega_m$ while the other parameters are like those of Fig.\ref{fig1}.}
	\label{fig2}
\end{figure}

Therefore, by having the value of $\omega_{sw}$, the appropriate value of $\omega_{c}$ to satisfy the condition $\Delta=\omega_m$ can be determined. As has been explained previously, for any fixed value of $\omega_{sw}$ the condition $\Delta=\omega_m$ leads to a third order algebraic equation which can give us a value for $\omega_c$ for which the system is stable. Therefore by fixing the coupling laser frequency at this specified value and by scanning the probe laser frequency around the effective cavity frequency, i.e., $\tilde{\omega}_0$, an experimental observation of a pattern like Fig.\ref{fig1} will be possible. 

On the other hand, in order to show how the pumping rate of the coupling laser affects the optical and atomic linear responses of the hybrid OMS to the external time-dependent perturbation, in Fig.\ref{fig2} we have plotted the normalized amplitudes of the anti-Stokes sideband $A_a=\omega_{R}|\tilde G^R_{aa^\dag}(\omega_{pc}/\omega_{R})|$ of the optical field \Big(Fig.\ref{fig2} (a)\Big), as well as the anti-Stokes sideband $A_c=\omega_{R}|\tilde G^R_{ca^\dag}(\omega_{pc}/\omega_{R})|$ of the Bogoliubov mode of the BEC \Big(Fig.\ref{fig2} (b)\Big), for three different values of the coupling laser pumping rate: $\eta_c=2.5\kappa$ (red solid curve),  $\eta_c=0.5\kappa$ (black dashed curve), and $\eta_c=7.5\kappa$ (blue dotted curve) while the \textit{s}-wave scattering frequency has been fixed at $\omega_{sw}=40\omega_{R}$ and the system is in the red detuned regime of $\Delta=\omega_{m}$. Since in the red detuned regime the Stokes sidebands are much weaker than Stokes ones and consequently the optical and atomic modes oscillate effectively with anti-stokes amplitudes, we no longer show the Stokes amplitudes in Fig.\ref{fig2}.

As is seen from Fig.\ref{fig2} (a), the position of the anti-resonance frequency of the optical field is invariant for different values of the coupling laser pumping rate since for the specified value of the \textit{s}-wave scattering frequency, i.e., $\omega_{sw}=40\omega_{R}$, the effective mechanical frequency of the Bogoliubov mode of the BEC is fixed. However, the amount of splitting between the normal modes of the optical and atomic fields becomes larger as the pumping rate of the coupling laser increases. It is because of the fact that the optical mean-field, i.e., $|\alpha|$, increases by increasing $\eta_c$ and consequently the enhanced effective optomechanical coupling between the optical and atomic fields, i.e, $\zeta|\alpha|$ increases. As is well-known in optomechanics \cite{Bowen book}, in the red detuned regime of $\Delta=\omega_m$, where the effective resonance frequencies of the two oscillators degenerate, the normal frequencies occur at  $\omega_{pc}\approx\omega_m\pm\zeta|\alpha|$. Therefore, it is obvious that by increasing the optical mean-field through the coupling laser pumping rate, the splitting between the two modes increases. In other words, the larger is the enhanced optomechanical coupling, the larger is the splitting between the normal modes \cite{Dobr,Grob}.

On the other hand, the same phenomenon of normal mode splitting also occurs for the Bogoliubov mode of the BEC where the amount of splitting increases by increasing the pumping rate of the coupling laser as is seen from Fig.\ref{fig2} (b). The difference is that at $\omega_{pc}=\tilde{\omega}_0$ where the phenomenon of the anti-resonance occurs for the optical field, the atomic filed amplitude reduces to a nonzero minimum.

\subsection{spectral function and effective damping rate of the cavity}
The response behavior of the optical mode to the external time-dependent perturbation can be explained in another way in terms of the sell-energy and the spectral function of the optical mode. The spectral function \cite{Coleman} which is usually interpreted as an effective density of single-particle states is defined as
\begin{equation}\label{SF}
\mathcal{A}(\omega)=-\frac{2}{\pi} {\rm{Im}} [\tilde G_{aa^\dag}^R(\omega)],
\end{equation}
for the optical field of the cavity. On the other hand, in order to obtain the optical self-energy \cite{optomechanicswithtwophonondriving}, it is enough to algebraically eliminate the Bogoliubov mode operators in the Fourier transform of the linearized QLEs of Eq.(\ref{nA}) so that  the optical field $a(\omega)$ can be written as
\begin{eqnarray}
-i\omega \delta a(\omega)&=& \! -\Big(i\Delta+\frac{\kappa}{2} + i\Sigma_a(\omega)\Big) \delta a(\omega)\nonumber\\
&& + \lambda_a (\omega) \delta a^\dag (\omega) + \!\! \sqrt{\kappa} A_{in}(\omega),\label{asolveOMS1 a}
\end{eqnarray}
where the optical self-energy is obtained as
\begin{equation} \label{Sigmaa}
\Sigma_a(\omega)=\frac{4|\alpha|^2\zeta^2(\omega_{sw}-2\Omega_c)}{\gamma^2-4i\gamma\omega-4\omega^2-\omega_{sw}^2},
\end{equation}
and $\lambda_a(\omega)=i\alpha\Sigma_a(\omega)/\alpha^{\ast}$. Furthermore, the last term in Eq.(\ref{asolveOMS1 a}) is the effective quantum noise operator which is a complex combination of the optical and atomic quantum noises. Since it has no role in our discussion we do not represent here its explicit expression.

As is seen from Eqs.(\ref{asolveOMS1 a}) and (\ref{Sigmaa}), the self-energy is a complex function of frequency whose real part modifies the frequency of the optical mode while its imaginary part modifies the damping rate of the cavity. In fact due to the interaction of the atomic mode with the optical field the damping rate of the cavity is modified through the imaginary part of the optical self-energy so that the effective damping rate of the cavity can be defined as
\begin{equation}\label{keff}
\kappa_{eff}(\omega)=\frac{\kappa}{2}-\rm Im[\Sigma_a(\omega)].
\end{equation}

\begin{figure}
	\centering
	\includegraphics[width=7cm]{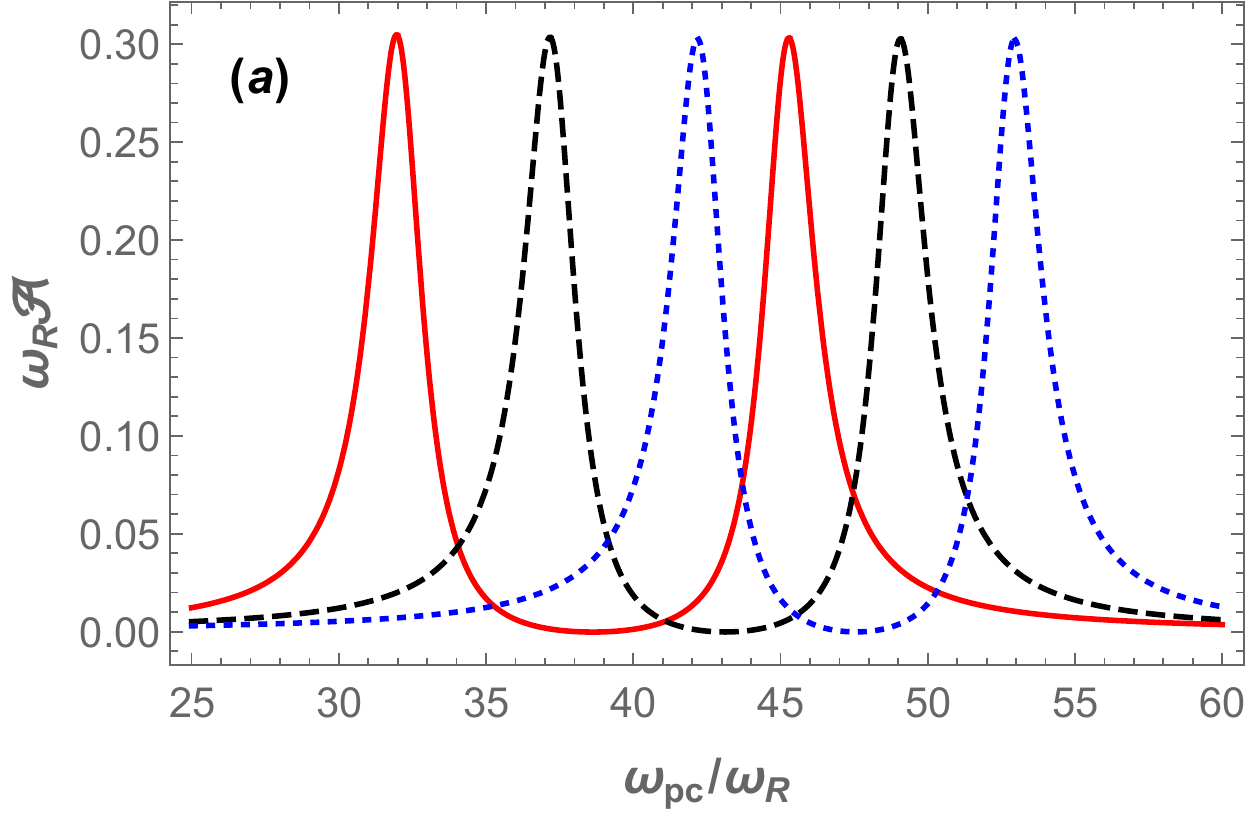}
	\includegraphics[width=7cm]{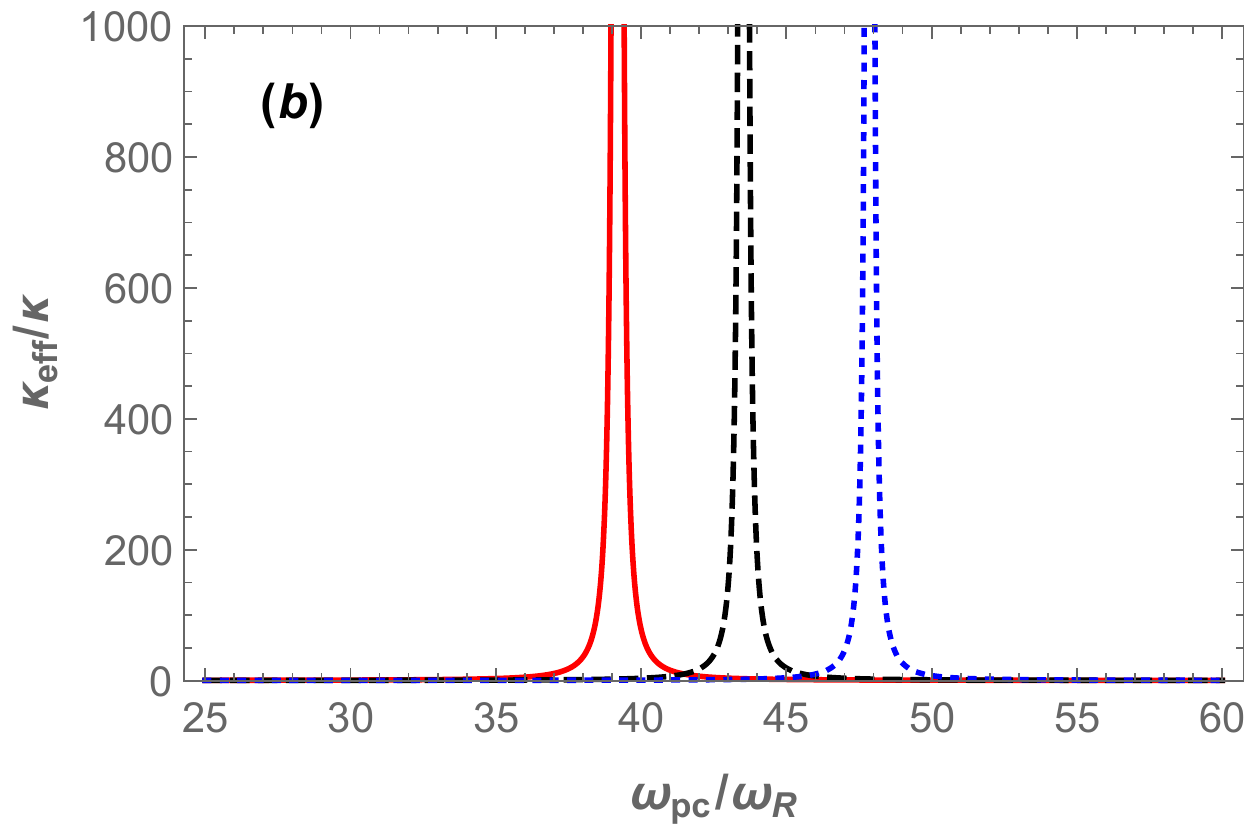}
	\caption{
		(Color online) (a) The normalized optical spectral function $\omega_R\mathcal{A}(\omega_{pc}/\omega_R)$ (b) the normalized effective damping rate of the cavity $\kappa_{eff}(\omega_{pc}/\omega_R)/\kappa$ for three different values of the \textit{s}-wave scattering frequency: $\omega_{sw}=40\omega_R$ (red solid curve), $\omega_{sw}=45\omega_R$ (black dashed curve), and $\omega_{sw}=50\omega_R$ (blue dotted curve). It has been assumed that $\Delta=\omega_m$ and $\eta_c=0.5\kappa$, and the other parameters are like those of Fig.\ref{fig1}.}
	\label{fig3}
\end{figure}

Now, using the optical spectral function of Eq.(\ref{SF}) and the effective damping rate of the cavity, i.e., Eq.(\ref{keff}), one can find an interpretation of the optical response behavior. For this purpose, in Fig.\ref{fig3} we have plotted the normalized optical spectral function $\omega_R\mathcal{A}(\omega_{pc}/\omega_R)$ \Big(Fig.\ref{fig3} (a)\Big) and the normalized effective damping rate of the cavity $\kappa_{eff}(\omega_{pc}/\omega_R)/\kappa$ \Big(Fig.\ref{fig3} (b)\Big) for three different values of the \textit{s}-wave scattering frequency: $\omega_{sw}=40\omega_R$ (red solid curve), $\omega_{sw}=45\omega_R$ (black dashed curve), and $\omega_{sw}=50\omega_R$ (blue dotted curve). It has been assumed that the system is in the red detuned regime of $\Delta=\omega_m$ while the cavity is driven at the rate of $\eta_c=0.5\kappa$ by the coupling laser.

A comparison between Figs.\ref{fig3} and \ref{fig1}(a) shows that at the two normal frequencies, where the system is at resonance and the optical anti-Stokes amplitude reaches to its peaks, i.e., at  $\omega_{pc}\approx\omega_m\pm\zeta|\alpha|$, the optical spectral function also maximizes for each value of $\omega_{sw}$ \Big(Fig.\ref{fig3} (a)\Big). On the other hand, the anti-resonance occurs at the frequency $\omega_{pc}=\omega_m$, where the optical spectral function becomes zero and the effective damping rate of the cavity reaches to a very sharp peak which is of the order of $10^6\kappa$, for each value of $\omega_{sw}$ \Big(Fig.\ref{fig3} (b)\Big). In this way, an interesting physical interpretation for the manifestation of two resonances and one anti-resonance arises.

It fact, at $\omega_{pc}=\omega_m$ where the effective damping rate of the cavity becomes very large and the optical spectral function is zero, it is expected that the oscillation amplitude of the optical mode goes to zero and the phenomenon of anti-resonance occurs while at the normal frequencies $\omega_{pc}=\omega_m\pm\zeta|\alpha|$ where the effective damping rate of the cavity becomes minimum but the optical spectral function maximizes, the optical mode oscillates with the maximum amplitude and the resonances occur.

\section{Conclusions}\label{Conclusions}
In conclusion, we have studied a hybrid OMS consisting of a cigar-shaped BEC which is affected by an external time-dependent perturbation. In the regime where the cavity photon number is low enough and under the Bogoliubov approximation, the BEC can be considered as a single mode quantum field which interacts with the cavity radiation pressure through an optomechanical coupling. In this way, the hybrid system behaves effectively as a standard ordinary OMS with the difference that there is an extra interaction in the system Hamiltonian which is due to the atomic collisions of the BEC atoms

Using the GLRT, which deals with the linear response of an open quantum system to an external time-dependent perturbation, we investigate the linear responses of the optical and atomic modes of the hybrid OMS while both the optical and the atomic modes are investigated as open quantum systems. The linear responses of the hybrid OMS are obtained through the solutions of the Green's functions equations of motion predicted by the GLRT. The great superiority of the GLRT over the SLRT is the fact that the dissipation is taken into consideration in the GLRT without necessity of any phenomenological manipulation. The main purpose of the present paper has been to show how some important phenomena like manifestation of the resonances an the anti-resonance in a hybrid OMS can be described based on a sophisticated theory which needs no phenomenological manipulations.

One of the most interesting features of the present hybrid OMS is that it behaves as a system consisting of of two coupled quantum oscillators which one of them (the Bogoliubov mode of the BEC) functions as an atomic parametric amplifier through the atom-atom interaction of the BEC atoms. The study of the linear responses of such hybrid OMS shows that it has two resonance frequencies corresponding to the two normal modes of the system and an anti-resonance frequency which occurs for the optical mode of the cavity which is driven directly by the time-dependent perturbation. It is demonstrated how the position of the anti-resonance frequency can be manipulated by the \textit{s}-wave scattering frequency of BEC atoms which itself is controllable through the transverse trapping frequency. Furthermore, it is also shown that the amount of splitting between the normal modes can be controlled by the coupling laser pumping rate which can change the effective optomechanical coupling between the optical and atomic modes. Finally, an interpretation of the optical response behavior especially the manifestation of the anti-resonance phenomenon is presented based on the optical spectral density and self-energy.

\end{document}